\RequirePackage{lineno}
\documentclass[twocolumn, prc,showpacs]{revtex4-1}
\usepackage{amsmath}
\usepackage{mathrsfs} 
\usepackage{amssymb} 
\usepackage{xcolor,graphicx}
\usepackage{subfigure}
\usepackage{dcolumn}
\usepackage{bm}
\usepackage{dashrule}
\usepackage[
  pdfstartview=FitH,
  bookmarksnumbered=true,
  bookmarksopen=true,
  colorlinks=true,
  pdfborder=002,
  citecolor=blue
]{hyperref}

\AtBeginDvi{} 
\usepackage{natbib}

\usepackage[latin1]{inputenc}

\begin{document}
  \newcommand {\nc} {\newcommand}
  \nc {\Sec} [1] {Sec.~\ref{#1}}
  \nc {\IR} [1] {\textcolor{red}{#1}}
  \nc {\IB} [1] {\textcolor{blue}{#1}}

\title{Deep Learning approaches for nuclear binding energy prediction: A comparative study of RNN, GRU and LSTM Models}

\author{Amir Jalili$^{1}$}
 \email[]{''Amir Jalili (Corresponding author)''}
 \email[]{jalili@zstu.edu.cn}
\author{Feng Pan$^{2,3}$}
\author{Ai Xi Chen$^{1}$}
\author{Jerry P. Draayer$^3$}
\affiliation{$^1$Zhejiang Key Laboratory of Quantum State Control and Optical Field Manipulation, Department of Physics, Zhejiang Sci-Tech University, Hangzhou 310018, China}
\affiliation{$^2$Department of Physics, Liaoning Normal University, Dalian 116029, P.R. China}
\affiliation{$^3$Department of Physics and Astronomy, Louisiana State University, Baton Rouge, LA 70803-4001, USA}


\begin{abstract}
This study investigates the application of deep learning models-recurrent neural networks, gated recurrent units, and long short-term memory networks-for predicting nuclear binding energies. Utilizing data from the Atomic Mass Evaluation (AME2020), we incorporate key nuclear structure features, including proton and neutron numbers, as well as additional terms from the liquid drop model and shell effects. Our comparative analysis demonstrates that the gated recurrent units model achieves the lowest root-mean-square error (\( \sigma_\mathrm{RMSE} \)) of 0.326 MeV, surpassing traditional regression-based approaches. To assess model reliability, we validate predictions using the Garvey-Kelson relations, obtaining an error of 0.202 MeV, and further test extrapolation capabilities using the WS, WS3, and WS4 models. The extrapolation analysis confirms the robustness of our approach, particularly in predicting binding energies for nuclei near the driplines. These results highlight the effectiveness of deep learning in nuclear BE predictions, highlighting its potential to enhance the accuracy and reliability of theoretical nuclear models.
\end{abstract}

\keywords{Recurrent neural networks (RNNs), Gated recurrent units (GRUs), Long short-term memory networks (LSTMs), Nuclear binding energy, Liquid drop model}

\maketitle

\section{Introduction}
\label{sec:intro}

The precise determination of nuclear binding energy (BE) is crucial for understanding the stability and structure of atomic nuclei, as well as the fundamental interactions that govern them. Traditional global models, such as the liquid drop model (LDM) \cite{ldm} and the Bethe-Weizsäcker (BW) formula \cite{ws1,ws2}, have long served as the foundation for nuclear mass predictions. While subsequent refinements, including the finite-range droplet model (FRDM) \cite{mol} and Weizsäcker-Skyrme (WS) model \cite{ws4}, have improved predictive accuracy, significant discrepancies persist, particularly near the nuclear drip lines.
Different kinds of nuclear mass models have been developed to incorporate more effects, which are known microscopic mass models based on the relativistic \cite{r1,r2,r3,r4,r5,r6,r7,r8,r9,r10,r11}, and nonrelativistic density functionals  with other theoretical models \cite{n1,n2,n3,n4,n5,n6,n7,n8,n9,n10}.
These challenges necessitate the exploration of alternative approaches that can effectively capture the complex nonlinear dependencies within nuclear data.
Theoretical models for nuclear mass predictions can generally be categorized into two main types: global models and local models \cite{l1,l2,l3,l4}. Global theoretical models, such as macroscopic-microscopic approaches and nuclear density functional theories \cite{pw}, aim to describe nuclear masses across the entire nuclear chart by utilizing a unified theoretical framework. While these models provide valuable insights into nuclear structure and large-scale trends, their accuracy may vary in regions where experimental data are scarce.

In contrast, local models rely on the assumption of strong correlations between neighboring nuclei and establish predictive relationships based on known experimental masses. These approaches, often referred to as local-type theoretical models, have distinct advantages in regions where experimental data are available, as they can achieve higher predictive accuracy by using well-established nuclear interactions. Among the most notable local models is the Garvey-Kelson (GK) relation \cite{gk1,gk2,gk3,gk4}, which exploits linear mass relationships between neighboring nuclei to estimate unknown masses with remarkable precision. Another widely used local approach is the neutron-proton interaction-based mass relation, which incorporates the effects of nucleon interactions to refine mass predictions.
Due to their reliance on empirical data, local models demonstrate exceptional accuracy within experimentally known regions. For instance, the GK relation and similar mass relations typically achieve an accuracy of approximately 0.2 MeV \cite{l3}, significantly outperforming many global models in these well-characterized regions. However, their extrapolation capabilities are inherently limited, making them less reliable for predicting masses far from experimentally measured nuclei. Consequently, the integration of both global and local models, along with advanced machine learning (ML) techniques, presents a promising avenue for enhancing the precision and robustness of nuclear mass predictions across the entire nuclear landscape.
In this paper, we will extend our analysis by testing and evaluating our  models using the GK relations, further assessing their reliability and accuracy in predicting nuclear BE.

In recent years, ML and deep learning (DL) techniques have emerged as powerful tools for nuclear physics applications, enabling accurate predictions of various nuclear properties.
In recent years, neural networks (NN) with various algorithms, along with advanced ML techniques, have been successfully applied to nuclear physics studies \cite{boe,wu}. Examples include convolutional NNs \cite{lu}, support vector machines \cite{yuk}, and Kolmogorov-Arnold networks \cite{ren}. Additionally, several NN-based models have been employed for predicting nuclear properties, such as binding energies \cite{liu,niu1,zhao,xie,niu2,gao,liu2,uta,lov,zeng,zhang,mum}, energy spectra \cite{he,wang2}, charge radii \cite{jalili1,dong,wu6}, $\alpha$-decay half-lives \cite{li,jalili2}, and $\beta$-decay half-lives \cite{jalili3,niu3}, among others \cite{wu7,shang1,shang2,za}.
Furthermore, kernel ridge regression has been applied for nuclear mass predictions, demonstrating its effectiveness in refining mass models and improving predictive accuracy \cite{wu1,wu2,wu5}. These advancements underscore the increasing role of ML in nuclear physics, providing robust and data-driven approaches for modeling complex nuclear phenomena.

Their ability to analyze complex patterns within large datasets has significantly enhanced our understanding of nuclear structure and decay processes, offering new insights beyond traditional theoretical models.
Unlike conventional models that rely on predefined functional forms, ML-based methods can extract intricate patterns from large datasets, enabling improved generalization and predictive performance. Among these, recurrent neural networks (RNNs) and their variants-gated recurrent units (GRUs) and long short-term memory (LSTM) networks-are particularly well-suited for sequential data modeling, making them promising candidates for nuclear properties predictions \cite{rnn1,rnn2,rnn3,rnn4,gru,lstm}. The advantage of RNNs stems from their feedback mechanism, which allows them to retain memory of previous inputs, effectively capturing long-range correlations in nuclear structure.

In this study, we employ RNN-based architectures to predict nuclear BE using data from the Atomic Mass Evaluation (AME2020) \cite{ame}. Our approach integrates nuclear structure features derived from the BW formula, including volume, surface, Coulomb, and pairing terms, alongside shell effects. By dynamically learning weights and feature interactions, our models surpass conventional regression-based approaches.
The results demonstrate that GRUs achieve the lowest \( \sigma_\mathrm{RMSE} \) of 0.326 MeV, significantly outperforming traditional models. To further validate our predictions, we apply the GK relations, which are highly accurate in regions where experimental data are available.  To evaluate the reliability of our approach, we embed the mass predictions from three different RNN models into the GK relations and obtain an error of approximately 0.202 MeV for the GRU case. This confirms the effectiveness of DL models in capturing nuclear mass trends with high precision.
Given that more than 11,000 nuclei require extrapolation, DL algorithms provide a powerful tool for extending nuclear BE predictions into unmeasured regions. Additionally, we investigate the extrapolation capabilities of our models in regions where experimental data are unavailable. Our study considers the WS \cite{ws}, WS3 \cite{ws3}, and WS4 \cite{ws4} models with both \texttt{relu} and \texttt{tanh} activation functions, ensuring the reliability of extrapolated predictions for unmeasured nuclei.

This paper is organized as follows: Section~\ref{sec:BW_RNN} presents the BW formula along with the fitting parameters and details of the RNN, LSTM, and GRU architectures, including data preprocessing and training procedures. Section~\ref{sec:results} discusses the results, covering the evaluation of RNN, GRU, and LSTM models, validation using GK relations, and an analysis of neutron and proton separation energies, mass excess, and extrapolation capabilities for the WS, WS3, and WS4  models. Finally, Section~\ref{sec:conclusion} concludes the study and outlines future directions for advancing DL-based nuclear binding energy models.

\section{Theory}\label{sec:BW_RNN}
\subsection{LDM}\label{21}

Traditional feed-forward NNs establish a direct, deterministic mapping between input features (\textbf{x}) and output (\textbf{y}) by optimizing complex non-linear functions \cite{vap0,jain}. In contrast, RNNs introduce an additional feedback mechanism, where the output of a neuron at a given time step is fed back into the network as an input for the subsequent time step \cite{b1,b2,b3,b4}. This feedback structure allows RNNs to effectively model sequential dependencies and long-range correlations within nuclear data, making them particularly suitable for nuclear BE predictions.

To construct our model, we employ features extracted from the BW mass formula \cite{ws2,kirson}. The LDM, which serves as the foundation for BW, describes the nucleus in terms of its fundamental constituents: protons ($Z$), neutrons ($N$), and atomic mass number ($A$). The total nuclear binding energy ($E_{\text{B}}$) is expressed as a sum of multiple energy contributions:

\begin{equation}
BE = a_\text{V} A - a_\text{S} A^{2/3} - a_\text{C} \frac{Z(Z - 1)}{A^{1/3}} - a_\text{A} \frac{(N - Z)^2}{A} + \delta(N, Z),
\end{equation}

where:
$a_\text{V} A$ represents the volume energy,
$a_\text{S} A^{2/3}$ accounts for the surface energy,
$a_\text{C} \frac{Z(Z - 1)}{A^{1/3}}$ denotes the Coulomb repulsion energy between protons,
$a_\text{A} \frac{(N - Z)^2}{A}$ corresponds to the symmetry energy associated with neutron-proton imbalance and
$\delta(N, Z)$ captures pairing effects and is given by:

\begin{equation}
\delta(N, Z) =
\begin{cases}
+\delta_0, & \text{for even } Z, N ~(\text{even } A), \\
0, & \text{for odd } A, \\
-\delta_0, & \text{for odd } Z, N ~(\text{even } A),
\end{cases}
\end{equation}
where $\delta_0 = a_\text{P} A^{k_{\rm P}}$.

Using a least squares fitting (LSF) approach, we determine the optimal values of the coefficients in our model. The resulting \( \sigma_\mathrm{RMSE} \) for a dataset comprising approximately 3264 nuclei (\(Z \geq 8, N \geq 8\)) is found to be 3.29 MeV. The fitted parameter values obtained through the LSF method are as follows:
$a_V = 15.45,
    a_S = 16.91,
    a_C = 0.70,
    a_A = 22.62,
    a_P = 9.35$ and
    $k_p = -\frac{3}{4}$.

\begin{figure}
  \includegraphics[height=8cm]{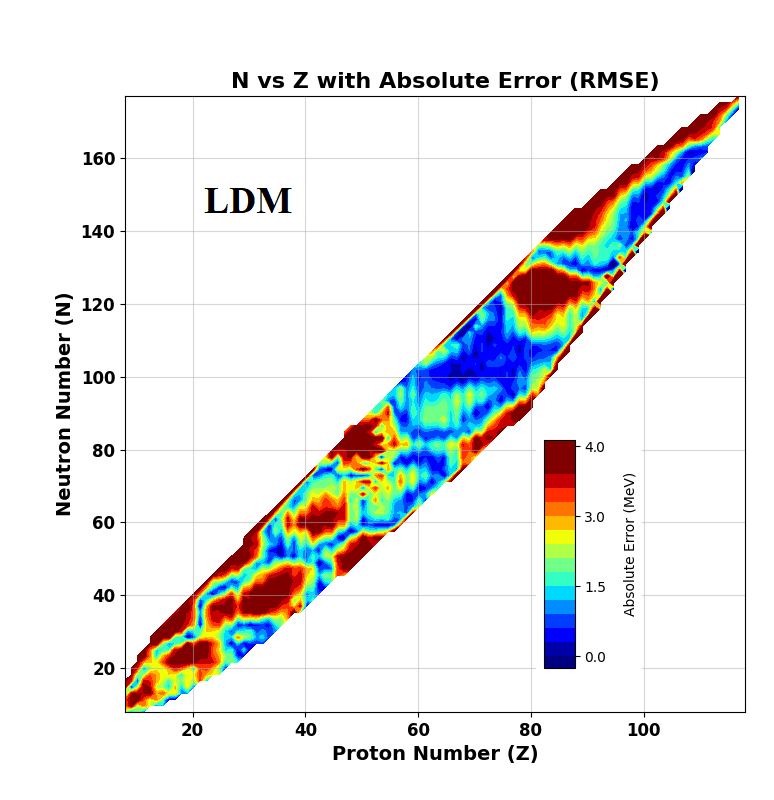}
\caption{Theoretical predictions for the total binding energy (BE) relative to experimental data from AME2020 for 3,092 nuclei.}\label{LDM}
\end{figure}

These values provide an optimized representation of nuclear BE contributions and serve as key inputs for refining predictive models.
Our objective is to apply the RNN architecture to refine these predictions by dynamically learning complex nuclear correlations, thereby reducing the deviation in BE estimations. The \( \sigma_\mathrm{RMSE} \) reduction achieved through our approach is presented in Fig.~\ref{LDM}.

\subsection{RNN, LSTM, and GRU Architectures}\label{3}
\subsubsection{Simple \textrm{RNN} Architecture}
\begin{figure}
  \includegraphics[height=8cm]{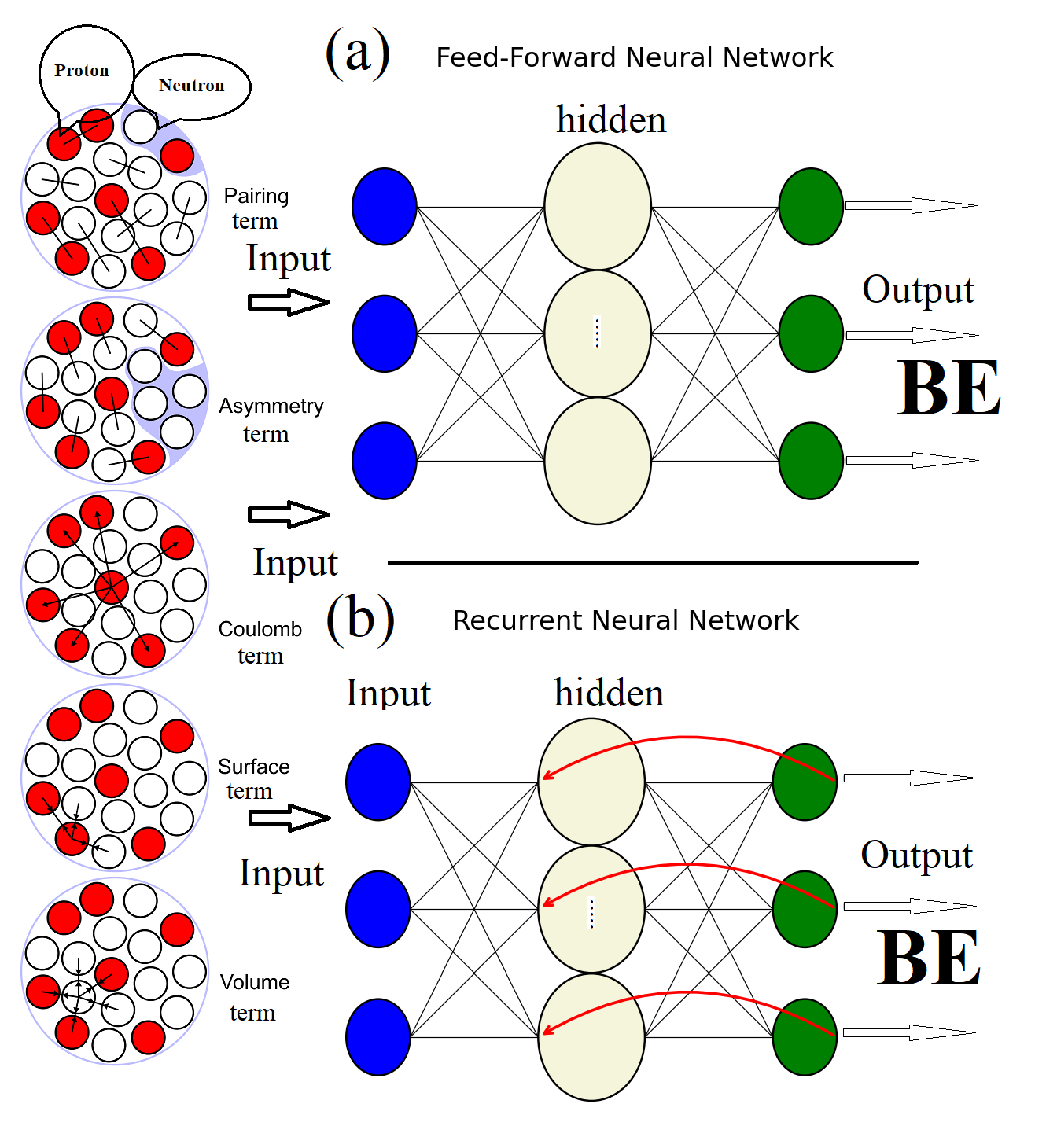}
\caption{Comparison between a Feed-Forward Neural Network (ANN), where inputs are processed in a unidirectional manner, and a RNN, which incorporates feedback connections by passing outputs of processing nodes back into the model to capture sequential dependencies.}\label{RNN}
\end{figure}

RNNs are a specialized class of NNs designed for processing sequential data, such as time series or feature sequences, through feedback connections \cite{b1}. Unlike traditional feedforward networks, which handle inputs independently, RNNs incorporate past outputs as additional inputs at each time step. This enables them to retain historical information, allowing RNNs to remember past states when making future predictions. As a result, they are particularly well-suited for tasks that require understanding temporal dependencies. See Fig.~\ref{RNN}

The concept of RNNs was first introduced by Rumelhart et al. (1986) \cite{b1} in a letter published in Nature, which described a self-organizing NN learning procedure. Over the years, RNNs have evolved into various forms, including input-output mapping networks, commonly used for classification and sequential data prediction. A major breakthrough in the field emerged  with the demonstration that RNNs can effectively perform credit assignment over long sequences, equivalent to processing information across 1,200 layers in an unfolded network. This advancement significantly improved the ability of RNNs to model complex sequential dependencies.  In 1997, one of the most influential architectures, LSTM, was introduced, enabling the processing of longer sequences by addressing the vanishing gradient problem \cite{b2,sc}.
\begin{figure*}
  \includegraphics[height=9cm]{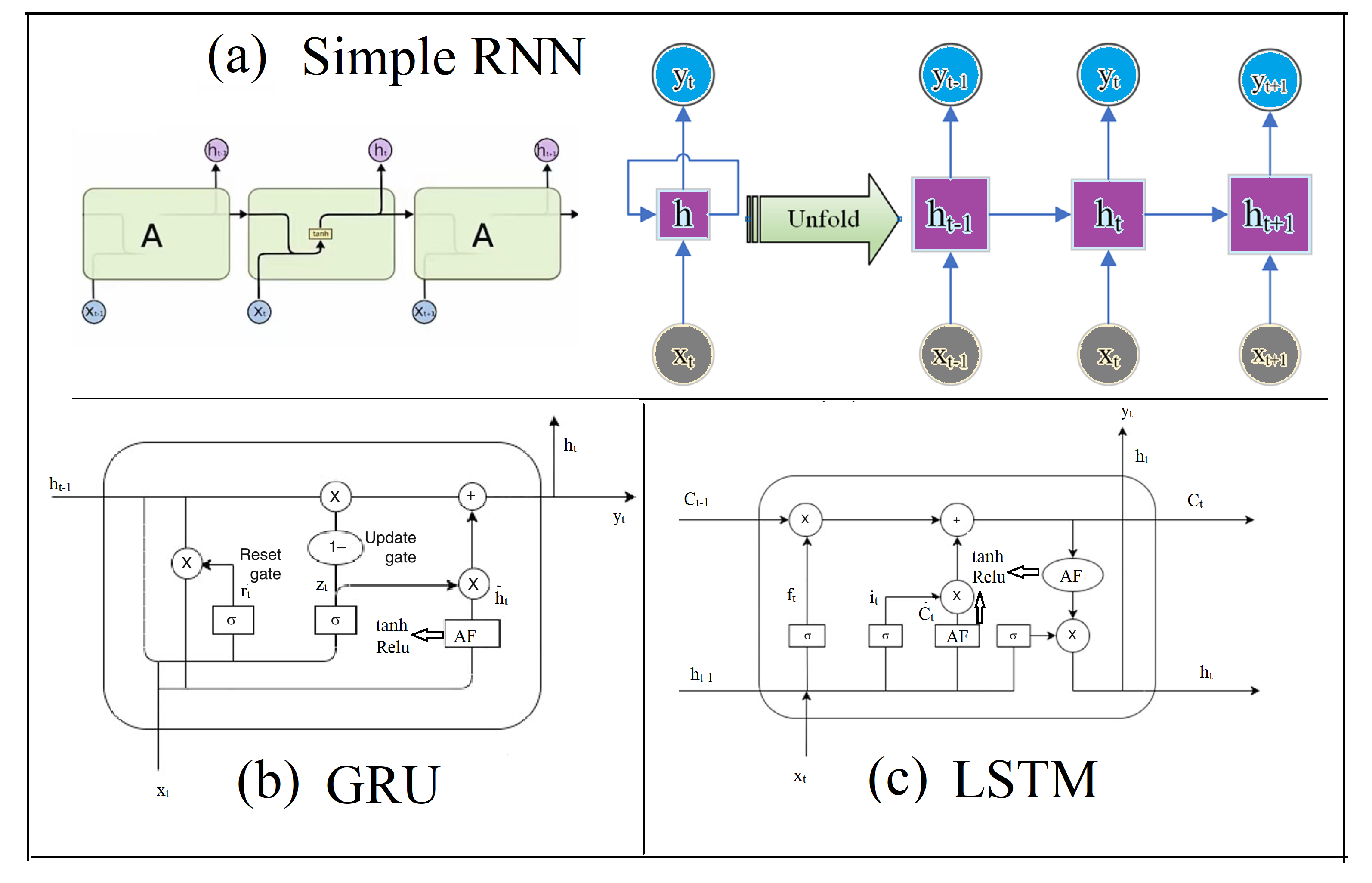}
 \caption{(a) Internal structure of a simple RNN cell, (b) architecture of a GRU cell, and (c) structure of an LSTM cell. Here, \( h_t \) represents the hidden state, and \( x_t \) denotes the input at time step \( t \). The forget gate (\( f_t \)), input gate (\( i_t \)), and output gate (\( y_t \)) regulate information flow in the LSTM model. \( C_t \) signifies the cell state at time step \( t \). AF refers to activation functions such as \textit{tanh} and \textit{relu}.}\label{Diag}
\end{figure*}
In this section, we introduce the three most prominent RNN architectures\textendash{}Simple RNN, LSTM, and GRU\textendash{}and highlight their significance in predicting nuclear binding energies.
The Simple RNN is the most fundamental recurrent architecture. It consists of a simple NN with a feedback connection that enables it to process sequential data of variable length. Unlike feedforward NNs, where each input feature has independent weights, RNNs share weights across multiple time steps, allowing efficient generalization over sequences.
In an RNN, the output at a given time step depends on previous time steps and is computed using a recursive update rule. This results in an unfolded computational graph, where weights are shared across time steps. Fig.~\ref{Diag} illustrates an RNN operating on an input sequence $x_t$ over \emph{t} time steps, where \emph{t} represents the position in the sequence rather than real-world time. The cyclical nature of the computational graph highlights how previous values influence the present step. The unfolded computational graph represents a chain of events, demonstrating the flow of information both forward (computing outputs and losses) and backward (computing gradients during backpropagation).

\textbf{Training RNNs for Nuclear Data Prediction}:

The training process for nuclear BE prediction involves computing gradients of the loss function concerning the model parameters. This consists of two key steps \cite{lip}:

1:Forward propagation: Information flows from left to right through the unfolded computational graph.

2:Backpropagation through time: The model iterates backward in time, computing gradients recursively for each node, starting from the final loss and propagating through all previous states.

Due to the sequential nature of forward propagation, gradient computations are computationally expensive, as parallelization is not feasible. To efficiently handle training, previously computed states from the forward pass are stored and reused in the backpropagation process. The total loss for a sequence is obtained by summing losses over all time steps. The output $o_t$ is passed through a softmax activation function, producing probability distributions over the predicted output categories.
One of the major challenges in training RNNs is the vanishing gradient problem, where gradients diminish exponentially as they are backpropagated through time. This leads to ineffective learning over long sequences. To mitigate this issue:
\texttt{tanh} is often used as the activation function, as it retains nonzero gradients for longer time steps.
We also explore the \texttt{relu} activation function to identify the best-performing configuration for nuclear BE predictions.

RNNs can be configured with different types of recurrent connections, affecting how information is transferred between layers:

Hidden-to-hidden connections: The RNN produces outputs at each time step, passing information between hidden units across time steps. This corresponds to the standard SimpleRNN architecture.
Output-to-hidden connections: Outputs at specific time steps are fed back into hidden units for future time steps, enhancing memory retention.
Sequential input to single output: The network processes an entire sequence before generating a single output, commonly used in applications like binding energy prediction.
This feature makes RNNs particularly well-suited for learning sequential relationships in nuclear structure data, where the BE exhibits correlations with proton and neutron numbers.

An RNN cell takes an input vector $x_t$ at time step $t$, processes it through a hidden state $h_t$, and generates an output $y_t$ using the following recurrence relations \cite{lip}:

\begin{equation}
    h_t = f(W_h x_t + U_h h_{t-1} + b_h)
\end{equation}

\begin{equation}
    y_t = f(W_y h_t + b_y)
\end{equation}

where:
- $W_h$, $U_h$, and $W_y$ are weight matrices controlling interactions between inputs, hidden states, and outputs.
- $b_h$ and $b_y$ are bias vectors.
- $f(\cdot)$ represents the activation function, typically a hyperbolic tangent \texttt{tanh} or \texttt{relu}.

The hidden state $h_t$ retains memory of past inputs, allowing RNNs to model nuclear binding energy as a function of historical nuclear configurations.

\subsubsection{\textrm{LSTM} Architecture}
Despite their advantages, RNNs suffer from the vanishing gradient problem, which hinders learning over long sequences. To address this, LSTM networks introduce gated memory cells that regulate information flow through input, forget, and output gates \cite{b2}. The LSTM equations are given by:

\begin{equation}
    f_t = \sigma(W_f x_t + U_f h_{t-1} + b_f)
\end{equation}

\begin{equation}
    i_t = \sigma(W_i x_t + U_i h_{t-1} + b_i)
\end{equation}

\begin{equation}
    \tilde{C}_t = \tanh(W_C x_t + U_C h_{t-1} + b_C)
\end{equation}

\begin{equation}
    C_t = f_t \odot C_{t-1} + i_t \odot \tilde{C}_t
\end{equation}

\begin{equation}
    y_t = \sigma(W_o x_t + U_o h_{t-1} + b_o)
\end{equation}

\begin{equation}
    h_t = o_t \odot \tanh(C_t)
\end{equation}

where:
- $f_t$, $i_t$, and $o_t$ are the forget, input, and output gates, respectively.
- $C_t$ is the cell state, maintaining long-term dependencies.
- $\sigma(\cdot)$ is the sigmoid function.
- The operator $\odot$ represents element-wise multiplication, ensuring that each gate influences only the corresponding elements of the cell state.

By incorporating nuclear structure features such as atomic mass ($A$), proton number ($Z$),  neutron number ($N$) and others into LSTM networks, we can dynamically learn interactions affecting nuclear BE and improve predictive accuracy.

\subsubsection{\textrm{GRU} Architecture}
A GRU is a simplified variant of LSTM that combines the forget and input gates into a single update gate, reducing computational complexity \cite{rez}. The GRU update equations are:

\begin{equation}
    r_t = \sigma(W_r x_t + U_r h_{t-1} + b_r)
\end{equation}

\begin{equation}
    z_t = \sigma(W_z x_t + U_z h_{t-1} + b_z)
\end{equation}

\begin{equation}
    \tilde{h}_t = \tanh(W_h (r_t \odot h_{t-1}) + U_h x_t + b_h)
\end{equation}

\begin{equation}
    h_t = z_t \odot \tilde{h}_t + (1 - z_t) \odot h_{t-1}
\end{equation}

where:
- $r_t$ is the reset gate, determining how much past information to forget.
- $z_t$ is the update gate, controlling the trade-off between past and new information.
- $h_t$ is the hidden state update.

GRUs provide a computationally efficient alternative to LSTMs while retaining memory for long-term dependencies in nuclear data. This allows GRUs to effectively capture binding energy variations across isotopic chains.

By training on binding energy datasets such as AME2020, our RNN, LSTM, and GRU models learn complex interactions between nuclear features, leading to improved generalization and predictive accuracy. The ability of these models to retain sequential dependencies is crucial for extrapolating BE values for unmeasured isotopes, reducing \( \sigma_\mathrm{RMSE} \) compared to traditional mass models.
In this work, we apply DL architectures, including RNNs, LSTMs, and GRUs, to enhance nuclear binding energy predictions. These models dynamically adjust their internal states based on sequential dependencies, making them highly effective for modeling complex nuclear interactions. The subsequent sections will outline our experimental setup, training methodology, and performance evaluation metrics.

\subsection{Nuclear features as  the training, validation and testing sets}\label{4}
The properties of an atomic nucleus are fundamentally determined by the number of protons and neutrons it contains, making them the most straightforward choice for input variables in predictive models. However, when dealing with limited datasets, incorporating additional engineered features\textendash{}beyond these fundamental parameters\textendash{}can significantly enhance the predictive power of NNs. These engineered features serve as priors that encode critical domain-specific information, a well-established practice in nuclear physics research (e.g., \cite{lov,mum,dong,dong2}).

For instance, Ref. \cite{dong2} demonstrated that supplementing  (\(Z\)) and  (\(N\)) numbers with two additional features representing pairing interactions and shell-closure effects substantially improved the accuracy of nuclear charge radius predictions compared to Bayesian models that relied solely on \(N\) and \(Z\). Similarly, in Ref. \cite{dong}, further enhancements were observed by incorporating two additional features that account for isospin dependence and local nuclear structure anomalies. These findings highlight the crucial role of feature engineering in improving nuclear BE predictions.
In our base model, denoted as RNN3, GRU3 and LSTM3, the input space includes the number of  (\(N\)), the number of (\(Z\)), and (\(A\)), with the sole prediction being the nuclear mass. For RNN7, GRU7 and LSTM7, we incorporate additional bulk properties, including the mass number (\(A\)), \(A^{2/3}\) (from volume and surface terms), \(Z(Z-1)/A^{1/3}\) from the Coulomb term, \((N-Z)^2/A\) from the asymmetry and $\delta_p=(-1)^Z+(-1)^N/2$ and incorporate information about magic numbers. RNN11, GRU11 and LSTM11, include pairing information, \(Z_{eo}\) and \(N_{eo}\), where \(Z_{eo}\) (\(N_{eo}\)) is 0 if \(Z\) (\(N\)) is even and 1 if \(Z\) (\(N\)) is odd,   and  ($V_N$) and ($V_Z$), representing number of valance nucleons. The features of the different NNs can be found in Table ~\ref{t1}.
To systematically evaluate the impact of feature selection and hyperparameter optimization, we designed three NN configurations with varying input layers. The first network utilizes three input features, the second expands to seven, and the final configuration includes eleven features, aligned with the terms of the LDM to assess deviations and their influence on BE predictions.

\begin{table}[h!]
\centering
\caption{Feature space in different RNNs structure.}\label{t1}
\label{t2}
\begin{tabular}{ccc}
    \hline
    \hline
    \textbf{Model} & \textbf{Input x} & \textbf{Output y} \\
    \hline
    \begin{tabular}{c}\textbf{RNN3}, \\ \textbf{GRU3}, \\ \textbf{LSTM3}\end{tabular} & $N$, $Z$, $A$ & BE  \\
    \hline
    \begin{tabular}{c}\textbf{RNN7}, \\ \textbf{GRU7}, \\ \textbf{LSTM7}\end{tabular} & \begin{tabular}{c}$N$, $Z$, $A$, $A^{2/3}$,  $Z(Z-1)/A^{1/3}$, \\ $(N-Z)^2/A$, $\delta_p$\\\end{tabular} & BE \\
    \hline
    \begin{tabular}{c}\textbf{RNN11}, \\ \textbf{GRU11}, \\ \textbf{LSTM11}\end{tabular} & \begin{tabular}{c}$N$, $Z$, $A$, $A^{2/3}$, $Z(Z-1)/A^{1/3}$, \\ $(N-Z)^2/A$, $\delta_p$,  $V_N$, $V_Z$, $Z_{eo}$, $N_{eo}$\end{tabular} & BE  \\
    \hline
    \hline
\end{tabular}
\end{table}

Given that our work involves a multidimensional feature space, we employed common techniques such as grid search and random search to fine-tune the hyperparameters, as detailed in Table III.

\begin{figure}
  \includegraphics[height=14cm]{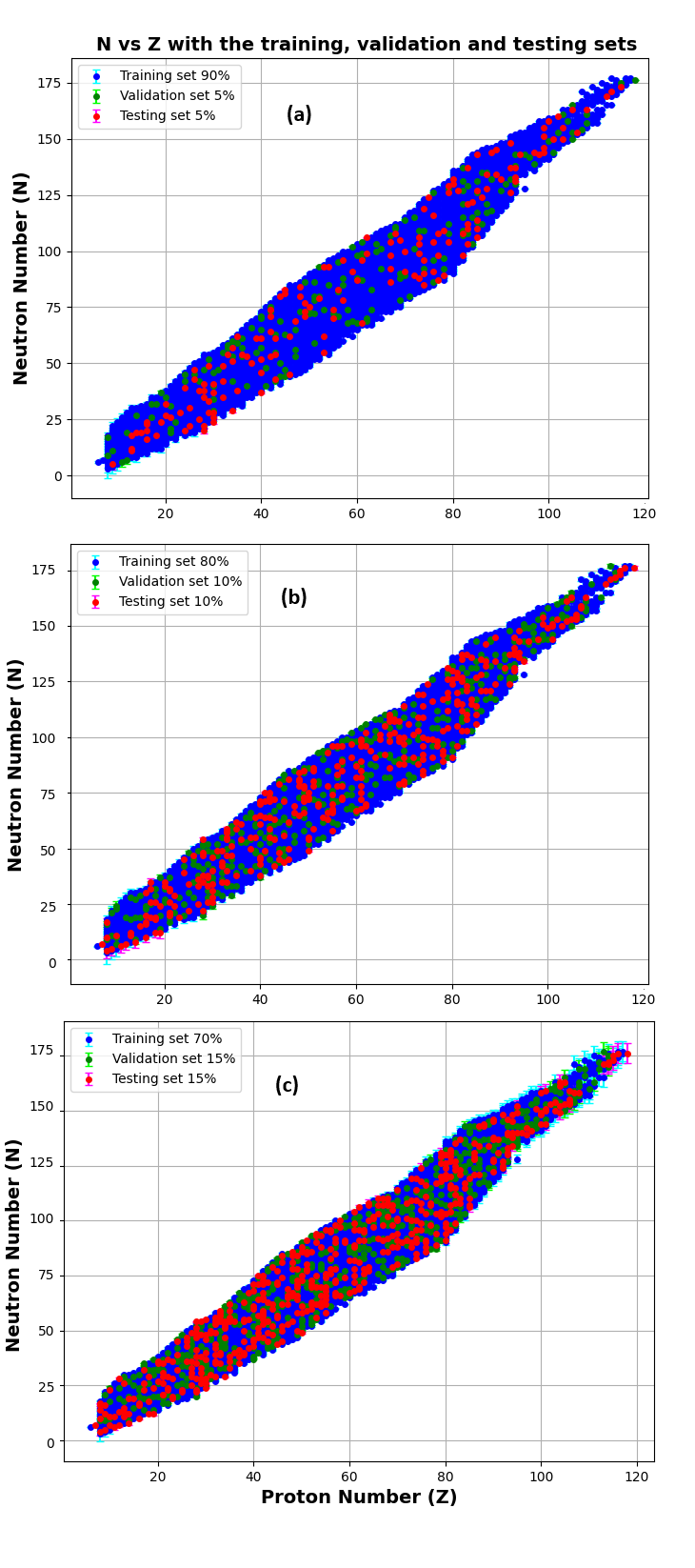}
\caption{Distribution of data among the training set (blue circles), validation set (green circles), and testing set (red circles) for RNN models with different splits: (a) 90\% training, 5\% validation, 5\% testing, (b) 80\% training, 10\% validation, 10\% testing, and (c) 70\% training, 15\% validation, 15\% testing. All data sets include nuclei from AME2020 \cite{ame}.}\label{tvs}
\end{figure}

Given the multitude of features, particularly in the domain of nuclear BE prediction using the LDM, where diverse inputs such as $N, Z, A$, and others are considered. The choice of feature combinations is contingent upon the characteristics of our data, such as the pairing term, Coulomb, or volume and surface terms. In our investigation, we systematically explored all possible combinations through a trial-and-error approach to determine the optimal values. To achieve this, we carefully prepared our training, validation and testing sets, spanning the nuclear landscape from $Z,N\geq8$  for absolute BE prediction, utilizing data from the AME2020.
In this implementation, we explore three types of RNN architectures. Each model is designed to capture temporal dependencies in the data effectively. The models are configured with specific numbers of units in their recurrent layers: 100 for SimpleRNN and LSTM, and 80 for GRU. The activation functions used are a combination of \texttt{relu} and \texttt{tanh}, which are chosen for their ability to handle the vanishing gradient problem and provide non-linearity. The dataset used for training these models consists of input features and target values, which are split into training, validation, and test sets with varying test sizes (0.1, 0.2, 0.3). The input data is normalized using the StandardScaler to ensure optimal performance during training.  Each model is compiled using the Adam optimizer and the mean squared error loss function, which is suitable for regression tasks. The training process involves feeding the scaled input data through the respective RNN layers, followed by a dense output layer that produces the final predictions. The performance of these models is evaluated based on the \( \sigma_\mathrm{RMSE} \) on the validation and test sets, providing insights into how well each architecture captures the underlying patterns in the data. This comparative analysis helps in understanding the strengths and limitations of each RNN variant for the given task.

\begin{table}[h]\label{t3}
    \centering
    \scriptsize
    \caption{The hyperparameter set of the RNN, LSTM, and GRU models.}
    \begin{tabular}{c c c}
        \hline
        \hline
        The $i^{th}$ layer & Layer Type & AF \\
        \hline
        0 & Input & - \\
        1 & Recurrent (RNN / LSTM / GRU) & \texttt{relu}, \texttt{tanh} \\
        2 & Fully Connected & \texttt{relu} \\
        3 & Output & Linear \\
        \hline
        \multicolumn{3}{l}{Other hyperparameters} \\
        \hline
        Hyperparameter & Value & Property \\
        \hline
        Batch size & 32 & - \\
        Objective function & MSE & Loss \\
        Optimizer & Adam & - \\
        Learning rate & $1 \times 10^{-3}$ & - \\
        \hline
        \hline
    \end{tabular}
    \label{tab:hyperparameters}
\end{table}


\subsection{BE data}\label{4}

The dataset used in this study is based on the AME2020 compilation \cite{ame}, which includes the BE of 3100 nuclei with proton and neutron numbers \(Z, N \geq 8\), encompassing both measured and extrapolated values. One of the key objectives of this work is to generate reliable predictions that can be valuable for future applications, particularly in nuclear astrophysics.

To ensure robust model evaluation, we employ multiple data partitioning strategies with varying training, validation, and testing splits: 90\%-5\%-5\%, 80\%-10\%-10\%,
and 70\%-15\%-15\% of the available data. In the first evaluation, we allocate 2,781 nuclei for training and 155 nuclei for testing. The second and third evaluations follow a different partitioning scheme. These different data splits allow us to assess the model's generalization capability under varying levels of training data availability. See Fig.~\ref{tvs}

Additionally, for validation and extrapolation, we incorporate extended datasets for different models \cite{ws,ws3,ws4}:
\begin{itemize}
    \item WS model: 11,248 nuclei
    \item WS3 model: 11,824 nuclei
    \item WS4 model: 11,879 nuclei
\end{itemize}

These expanded datasets enable further evaluation of model performance beyond the training regime, providing insights into their predictive reliability across a broader range of nuclear masses.

\section{Results and Discussion}\label{sec:results}
\subsection{Evaluation of RNN, GRU and LSTM}

To quantify how well the RNNs can describe nuclear BEs in the training, validation, and testing sets for different activation functions and learning rates, we use the standard deviation
\( \sigma_\mathrm{RMSE} \) between the predicted RNN values, \( BE ^\mathrm{RNNs} \), and the experimental data from AME2020, defined as

\begin{equation}
\sigma_\mathrm{RMSE} = \sqrt{\frac{\sum \limits _{i=1}^{\nu} (BE ^\mathrm{Exp}_i -BE ^\mathrm{RNNs}_i)^2}{\nu}},
\end{equation}
where \( \nu \) is the total number of nuclei considered.

In this phase of our study, we systematically expanded the input space by introducing new terms based on the BW mass formula. This augmentation aims to enhance the predictive capabilities of our RNN models. Tables III, IV, and V provide a comprehensive evaluation of different test sizes employed across various RNN architectures.

\begin{figure}
  \includegraphics[height=14cm]{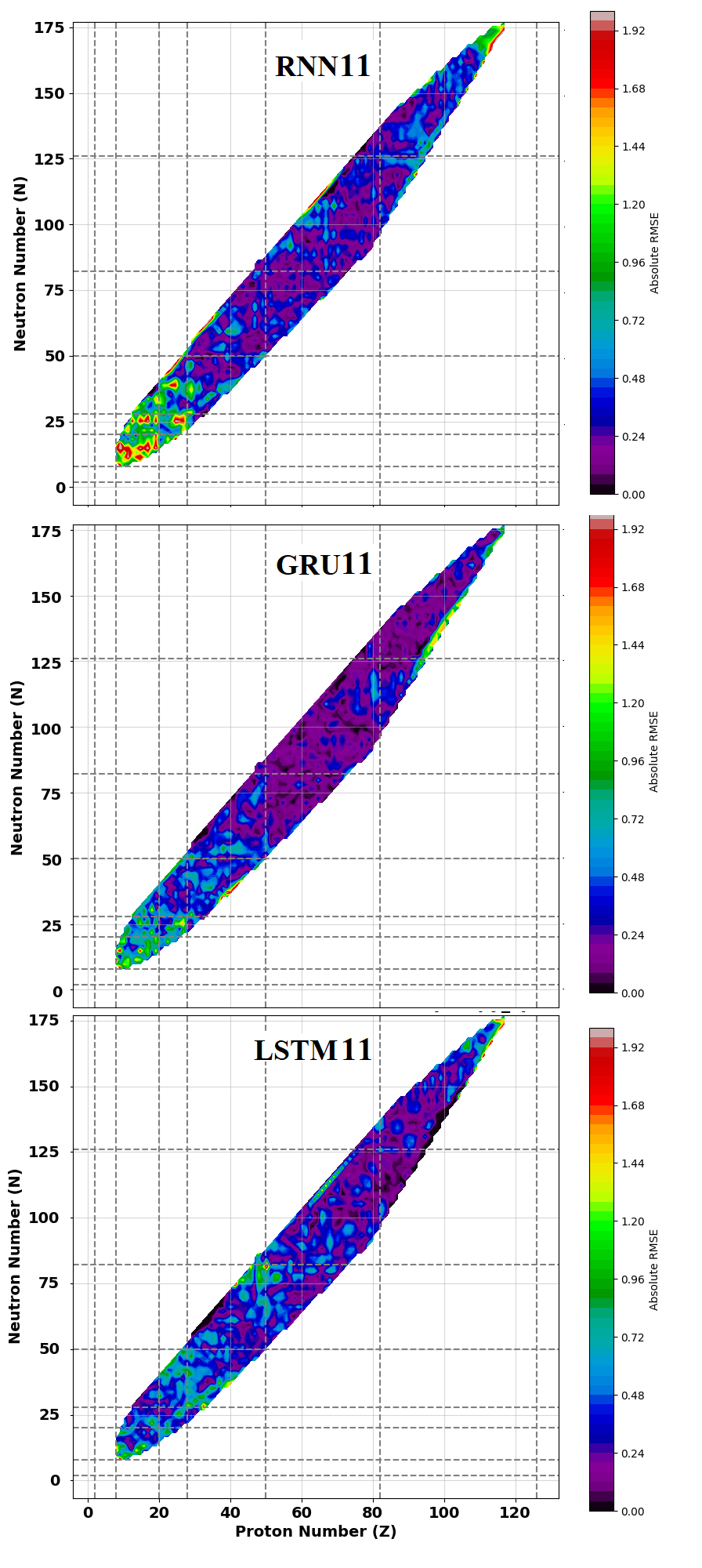}
  \caption{The absolute $\sigma_\mathrm{RMSE}$ value of BE between RNNs predictions using RNN11, GRU11 and LSTM11 features (see Table V) for 90\%,5\%,5\% . The $\sigma_\mathrm{RMSE}$ for all the models are also provided in Table V.}\label{error}
\end{figure}

\begin{table}[ht]\label{t3}
\centering
\caption{Comparison of training, validation, and testing $\sigma_{\textrm{RMSE}}$ for RNN3, GRU3, and LSTM3 across different learning rates, model architectures, and activation functions.}
\begin{tabular}{cccccc}
\hline
\textbf{LR} & \textbf{Model} & \textbf{AF} & \textbf{Training} & \textbf{Validation} & \textbf{Testing} \\
\hline
90\%,5\%,5\% & RNN3 & \texttt{tanh} & 1.1584 & 1.1289 & 1.1239 \\
90\%,5\%,5\%& RNN3 & \texttt{relu} & 1.3270 & 1.6657 & 1.3647 \\
90\%,5\%,5\% & GRU3 & \texttt{tanh} & 1.0673 & 1.0419 & 1.1476 \\
90\%,5\%,5\% & GRU3 & \texttt{relu} & 1.3844 & 1.2734 & 1.4251 \\
90\%,5\%,5\% & LSTM3 & \texttt{tanh} & 1.1624 & 1.1299 & 1.2123 \\
90\%,5\%,5\% & LSTM3 & \texttt{relu} & 1.2855 & 1.1604 & 1.2555 \\
80\%,10\%,10\% & RNN3 & \texttt{tanh} & 1.2281 & 1.2440 & 1.1551 \\
80\%,10\%,10\% & RNN3 & \texttt{relu} & 1.3022 & 1.2619 & 1.4610 \\
80\%,10\%,10\% & GRU3 & \texttt{tanh} & 1.1056 & 1.1652 & 1.0758 \\
80\%,10\%,10\% & GRU3 & \texttt{relu} & 1.3304 & 1.3929 & 1.2802 \\
80\%,10\%,10\% & LSTM3 & \texttt{tanh} & 1.1861 & 1.2573 & 1.1869 \\
80\%,10\%,10\% & LSTM3 & \texttt{relu} & 1.5782 & 1.6672 & 1.6093 \\
70\%,15\%,15\% & RNN3 & \texttt{tanh} & 1.2105 & 1.1831 & 1.2231 \\
70\%,15\%,15\% & RNN3 & \texttt{relu} & 1.7970 & 1.6499 & 1.6783 \\
70\%,15\%,15\% & GRU3 & \texttt{tanh} & 1.0838 & 1.0930 & 1.0192 \\
70\%,15\%,15\% & GRU3 & \texttt{relu} & 1.3233 & 1.3304 & 1.3790 \\
70\%,15\%,15\% & LSTM3 & \texttt{tanh} & 1.2303 & 1.3141 & 1.2976 \\
70\%,15\%,15\% & LSTM3 & \texttt{relu} & 1.4853 & 1.3755 & 1.5193 \\
\hline
\end{tabular}
\end{table}

\begin{table}[ht]\label{t4}
\centering
\caption{Comparison of training, validation, and testing $\sigma_{\textrm{RMSE}}$ for RNN7, GRU7, and LSTM7 across different learning rates, model architectures, and activation functions.}
\begin{tabular}{cccccc}
\hline
\textbf{LR} & \textbf{Model} & \textbf{AF} & \textbf{Training} & \textbf{Validation} & \textbf{Testing} \\
\hline
90\%,5\%,5\% & RNN7 & \texttt{tanh} & 0.5912 & 0.6490 & 0.6594 \\
90\%,5\%,5\% & RNN7 & \texttt{relu} & 0.7541 & 0.8457 & 0.8737 \\
90\%,5\%,5\% & GRU7 & \texttt{tanh} & 0.4447 & 0.4239 & 0.5473 \\
90\%,5\%,5\% & GRU7 & \texttt{relu} & 0.6231 & 0.7203 & 0.7988 \\
90\%,5\%,5\% & LSTM7 & \texttt{tanh} & 0.6044 & 0.5541 & 0.6492 \\
90\%,5\%,5\% & LSTM7 & \texttt{relu} & 0.8475 & 0.8677 & 1.0428 \\
80\%,10\%,10\% & RNN7 & \texttt{tanh} & 0.6633 & 0.7220 & 0.7431 \\
80\%,10\%,10\% & RNN7 & \texttt{relu} & 0.7897 & 0.8097 & 0.8413 \\
80\%,10\%,10\% & GRU7 & \texttt{tanh} & 0.5424 & 0.5673 & 0.5681 \\
80\%,10\%,10\% & GRU7 & \texttt{relu} & 0.7476 & 0.7817 & 0.7998 \\
80\%,10\%,10\% & LSTM7 & \texttt{tanh} & 0.5390 & 0.6568 & 0.6434 \\
80\%,10\%,10\% & LSTM7 & \texttt{relu} & 0.9369 & 1.1170 & 1.1298 \\
70\%,15\%,15\% & RNN7 & \texttt{tanh} & 0.7528 & 0.7999 & 0.7260 \\
70\%,15\%,15\% & RNN7 & \texttt{relu} & 0.7791 & 0.9288 & 0.8483 \\
70\%,15\%,15\% & GRU7 & \texttt{tanh} & 0.5372 & 0.6290 & 0.6152 \\
70\%,15\%,15\% & GRU7 & \texttt{relu} & 0.7914 & 0.9425 & 0.8951 \\
70\%,15\%,15\%3 & LSTM7 & \texttt{tanh} & 0.5138 & 0.6363 & 0.6026 \\
70\%,15\%,15\%3 & LSTM7 & \texttt{relu} & 0.9533 & 1.0979 & 1.0413 \\
\hline
\end{tabular}
\end{table}

\begin{table}[ht]\label{t5}
\centering
\caption{Comparison of training, validation, and testing $\sigma_{\textrm{RMSE}}$ for RNN11, GRU11, and LSTM11 across different learning rates, model architectures, and activation functions.}
\begin{tabular}{cccccc}
\hline
\textbf{LR} & \textbf{Model} & \textbf{AF} & \textbf{Training} & \textbf{Validation} & \textbf{Testing} \\
\hline
90\%,5\%,5\% & RNN11 & \texttt{tanh} & 0.4931 & 0.5022 & 0.6016 \\
90\%,5\%,5\% & RNN11 & \texttt{relu} & 0.5030 & 0.5388 & 0.5141 \\
90\%,5\%,5\% & GRU11 & \texttt{tanh} & 0.3294 & 0.3791 & 0.4591 \\
90\%,5\%,5\% & GRU11 & \texttt{relu} & 0.5618 & 0.5518 & 0.5632 \\
90\%,5\%,5\% & LSTM11 & \texttt{tanh} & 0.4817 & 0.5537 & 0.5574 \\
90\%,5\%,5\% & LSTM11 & \texttt{relu} & 0.4942 & 0.6965 & 0.6074 \\
80\%,10\%,10\% & RNN11 & \texttt{tanh} & 0.5306 & 0.5521 & 0.6023 \\
80\%,10\%,10\% & RNN11 & \texttt{relu} & 0.5707 & 0.5667 & 0.5018 \\
80\%,10\%,10\% & GRU11 & \texttt{tanh} & 0.3975 & 0.5045 & 0.4332 \\
80\%,10\%,10\% & GRU11 & \texttt{relu} & 0.4617 & 0.6180 & 0.5687 \\
80\%,10\%,10\% & LSTM11 & \texttt{tanh} & 0.5075 & 0.5753 & 0.5503 \\
80\%,10\%,10\% & LSTM11 & \texttt{relu} & 0.5326 & 0.7173 & 0.7356 \\
70\%,15\%,15\% & RNN11 & \texttt{tanh} & 0.4331 & 0.5071 & 0.5140 \\
70\%,15\%,15\% & RNN11 & \texttt{relu} & 0.5260 & 0.5744 & 0.5150 \\
70\%,15\%,15\% & GRU11 & \texttt{tanh} & 0.4057 & 0.5084 & 0.5258 \\
70\%,15\%,15\% & GRU11 & \texttt{relu} & 0.5278 & 0.5233 & 0.5272 \\
70\%,15\%,15\% & LSTM11 & \texttt{tanh} & 0.4798 & 0.5300 & 0.5791 \\
70\%,15\%,15\% & LSTM11 & \texttt{relu} & 0.4700 & 0.6147 & 0.6969 \\
\hline
\end{tabular}
\end{table}

A key observation from the results is the significant deviation in $\sigma_\mathrm{RMSE}$ for certain cases. For instance, in Table III, the GRU3 model with the \texttt{relu} activation function exhibits a notable deviation in the testing set ($\sim 1.42$ MeV), which is higher compared to the \texttt{tanh} counterpart ($\sim 1.15$ MeV). Similarly, LSTM3 with \texttt{relu} activation also demonstrates relatively poor performance, reaching an \( \sigma_\mathrm{RMSE} \) above $1.6$ MeV in some cases.
A substantial improvement is observed in Table IV, where RNN7 models incorporating surface and Coulomb terms lead to an approximate $50\%$ reduction in error compared to RNN3. Notably, GRU7 with \texttt{tanh} activation achieves an \( \sigma_\mathrm{RMSE} \) of $0.54$ MeV in the testing set, contrasting with the \texttt{relu} activation case, which remains closer to $0.80$ MeV.
Moving to Table V, the most striking improvement is evident in GRU11 models. The transition from GRU3 to GRU11 results in a $68.9\%$ reduction in \( \sigma_\mathrm{RMSE} \), confirming the critical role of feature expansion. GRU11 with \texttt{tanh} activation exhibits the lowest deviation, with training devation values around $0.326$ MeV, whereas some RNN11 cases still hover near $0.50$ MeV.

Overall, our findings suggest that GRU models outperform both standard RNN and LSTM architectures, particularly when employing the \texttt{tanh} activation function. The progressive enhancement from RNN3, GRU3, and LSTM3 to their respective 11-feature variants underscores the necessity of incorporating detailed nuclear properties for more accurate binding energy predictions.

Fig.~\ref{error} presents a comparative analysis of three different RNN architectures\textendash{}RNN11, GRU11, and LSTM11\textendash{}each utilizing 11 input features with the \(\tanh\) activation function. The color gradient represents the absolute \( \sigma_\mathrm{RMSE} \) in MeV, with darker shades indicating lower deviations and brighter colors (yellow-red) highlighting regions of higher error.

The error distribution shows a clear trend where deviations are more pronounced for nuclei with low proton (\(Z\)) and neutron (\(N\)) numbers. This is particularly evident from the red and yellow regions in the lower left of each subplot.
This suggests that the models struggle more with light nuclei, likely due to stronger shell effects and nuclear structure complexities that are not fully captured by the input features.
The intricate nature of  magic numbers contributes to the sophisticated trends observed, underscoring the complex interplay between model features and the representation of nuclear phenomena.
However, it is crucial to admit that some deviations are observed for all cases in light mass nuclei, specifically those near to $Z, N$=8,20.
Regarding the performance across architectures, the GRU11 model exhibits the best overall performance, as indicated by a larger fraction of the domain covered in dark blue to purple shades, which correspond to lower \( \sigma_\mathrm{RMSE} \) values.
The RNN11 model shows more scattered high-error regions, especially in the light-nuclei regime and near the neutron-rich dripline.
So, the LSTM11 model also performs well but exhibits slightly larger deviations compared to GRU11, particularly at the neutron-rich and proton-rich extremes.
The GRU11 model achieves the lowest error across most of the nuclear chart, making it the most reliable among the tested architectures.
\begin{figure}
  \includegraphics[height=14cm]{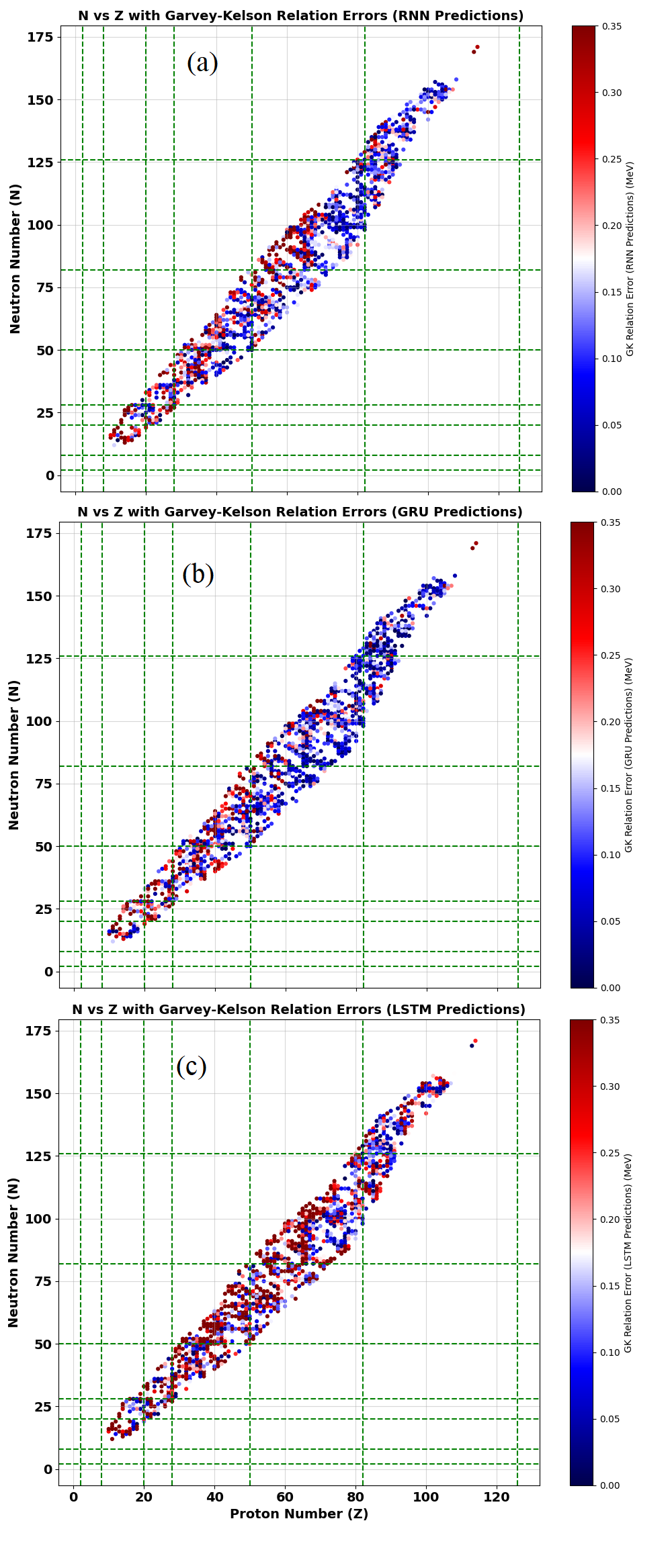}
\caption{Absolute deviations (in MeV) between the GK mass relations and the predicted masses using the RNN11, GRU11, and LSTM11 models.}\label{GK}
\end{figure}

\subsection{Evaluation with Garvey-Kelson relations}
In this section, we emphasize the potential application of the GK mass relations \cite{gk1} to validate and further analyze the BE errors of our NN models. The GK relations are known for their exceptional predictive power in nuclear mass evaluations, often yielding errors significantly lower than those found in conventional macroscopic-microscopic or purely microscopic mass formulae.
The GK relations for nuclear masses are defined as follows:

For nuclei with $N \geq Z$:
\begin{align}\label{eq:GK1}
    &M(Z-2, N+2) - M(Z, N) \nonumber \\
    &+ M(Z-1, N) - M(Z-2, N+1) \nonumber \\
    &+ M(Z, N+1) - M(Z-1, N+2) \approx 0,
\end{align}

and for nuclei with $Z < N$, the GK relation is given by:
\begin{align}\label{eq:GK2}
    &M(Z+2, N-2) - M(Z, N) \nonumber \\
    &+ M(Z, N-1) - M(Z+1, N-2) \nonumber \\
    &+ M(Z+1, N) - M(Z+2, N-1) \approx 0.
 \end{align}
Here, $M(Z, N)$ represents the nuclear mass for a nucleus with proton number $Z$ and neutron number $N$.

To assess the reliability of our best-performing RNN models, we compare their predictions against the GK relations. Specifically, we evaluate the deviations of RNN11, GRU11, and LSTM11-each employing 11 input features with the \(\tanh\) activation function-relative to the GK-predicted masses.
The results indicate that the deviation of the GK relations from the GRU11 model is approximately 0.202 MeV, which is notably lower than the 0.327 MeV deviation obtained from LDM-based features using the \(\tanh\) activation function. Similarly, the deviation for RNN11 with the GK relation is found to be 0.203 MeV, while LSTM11 exhibits a slightly higher deviation of 0.348 MeV.

The mean absolute error (MAE) for a given model is computed as follows:

\[
\text{MAE} = \frac{1}{N} \sum_{i=1}^{N} \left| M_i^{\text{pred}} - M_i^{\text{exp}} \right|
\]

where \(M_i^{\text{pred}}\) and \(M_i^{\text{exp}}\) denote the predicted and experimental nuclear masses, respectively, and \(N\) represents the number of nuclei in the dataset.

Table \ref{tab:gk_errors} presents the \( \sigma_\mathrm{RMSE} \) and MAE values obtained for training, validation, and testing phases of the RNN11, GRU11, and LSTM11 models, alongside the \( \sigma_\mathrm{RMSE} \) of the GK relation errors for each model.

\begin{table}[h]
    \centering
    \caption{Performance metrics of RNN11, GRU11, and LSTM11 models evaluated against the GK relations.}
    \label{tab:gk_errors}
    \begin{tabular}{lcccc}
        \hline
        Model & \( \sigma_\mathrm{RMSE} \) (MeV) & MAE (MeV) & \( \sigma_\mathrm{RMSE} \) (GK Errors) \\
        \hline
        \textbf{RNN11} & & & \\
        Training   & 0.564  & 0.412 & 0.203 \\
        Validation & 0.596  & 0.429 & 0.223 \\
        Testing    & 0.725  & 0.480 & 0.239 \\
        \hline
        \textbf{GRU11} & & & \\
        Training   & 0.506  & 0.358 & 0.202 \\
        Validation & 0.546  & 0.393 & 0.209 \\
        Testing    & 0.624  & 0.412 & 0.228 \\
        \hline
        \textbf{LSTM11} & & & \\
        Training   & 0.571  & 0.437 & 0.348 \\
        Validation & 0.705  & 0.507 & 0.342 \\
        Testing    & 0.681  & 0.499 & 0.332 \\
        \hline
    \end{tabular}
\end{table}

These findings affirm that the GK relations serve as a robust benchmark, substantiating the reliability of our RNN-based mass predictions. The strong agreement, particularly with the GRU11 model, underscores the capability of NNs to capture nuclear mass systematics with high fidelity. However, further investigations are necessary to assess the model's generalization ability in extrapolation regions, such as near the nuclear drip lines, where experimental data are sparse. See Fig.~\ref{GK}

\begin{figure*}
  \includegraphics[height=6.5cm]{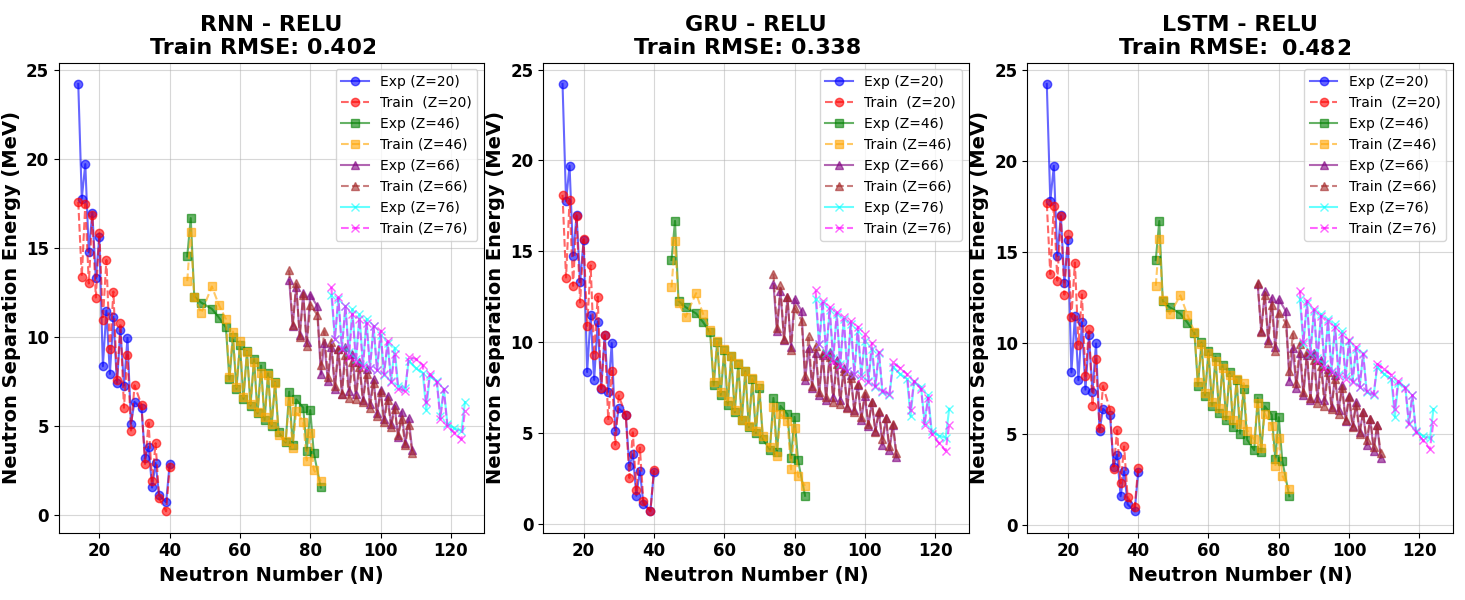}
\caption{Comparison of experimental single neutron separation energies with predictions from RNN11, GRU11, and LSTM11 models for $_{20}$Ca, $_{46}$Pd, $_{66}$Dy, and $_{76}$Os isotopes. All (\(\sigma_\mathrm{RMSE}\)) are given in MeV.}\label{SN}
\end{figure*}
\begin{figure*}
  \includegraphics[height=6.5cm]{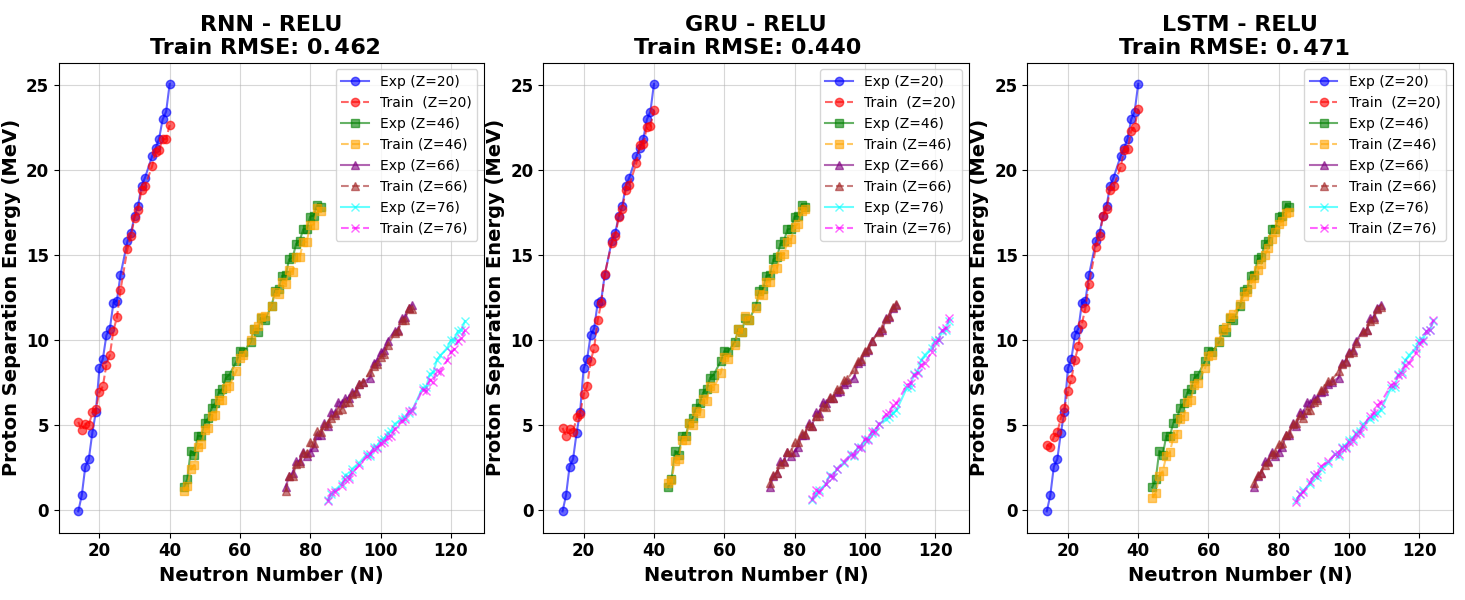}
  \caption{Comparison of experimental single proton separation energies with predictions from RNN11, GRU11, and LSTM11 models for $_{20}$Ca, $_{46}$Pd, $_{66}$Dy, and $_{76}$Os isotopes. All (\(\sigma_\mathrm{RMSE}\)) are given in MeV.}\label{SP}
\end{figure*}
\begin{figure*}
  \includegraphics[height=6.5cm]{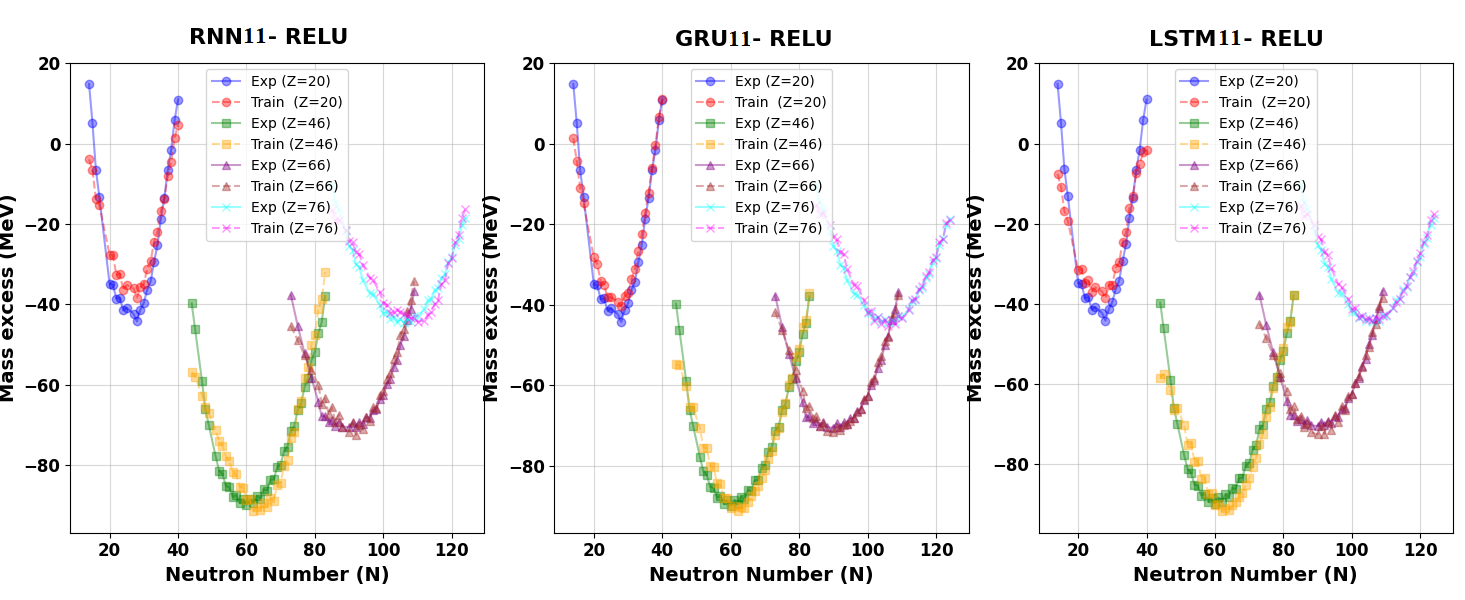}
  \caption{Comparison of experimental mass excess energies with predictions from RNN11, GRU11, and LSTM11 models for $_{20}$Ca, $_{46}$Pd, $_{66}$Dy, and $_{76}$Os isotopes.}\label{ME}
\end{figure*}

\subsection{Evaluation of Neutron and Proton Separation Energies and Mass Excess}
In pursuit of reducing the \( \sigma_\mathrm{RMSE} \) discrepancy between RNNs predictions and AME2020, a secondary objective is to navigate specific mass regions.
This distinction is particularly evident in the \( \sigma_\mathrm{RMSE} \) variations, which represent the absolute differences in neutron separation energy (\( S_n \)) and proton separation energy (\( S_p \)) between the predictions from different RNN models and the experimental data from AME2020.

The single-neutron and single-proton separation energy curves further reinforce the importance of these terms \cite{zeng}:

\begin{equation}
    M(Z, N) = Z M_p + N M_n - B(Z, N)
\end{equation}

\begin{equation}
    S_n(Z, N) = B(Z, N) - B(Z, N-1)
\end{equation}

\begin{equation}
    S_p(Z, N) = B(Z, N) - B(Z-1, N)
\end{equation}

where \( M(Z, N) \) represents the nuclear mass, which is related to the total binding energy \( B(Z, N) \), and \( M_n \) and \( M_p \) denote the neutron and proton rest masses, respectively.
Figs.~\ref{SN},~\ref{SP} and ~\ref{ME} compare the neutron and proton separation energies, as well as mass excess, for isotopic chains of four different elements: $_{20}$Ca, $_{46}$Pd, $_{66}$Dy, and $_{76}$Os isotopes. The absence of shell effects and pairing terms in the RNN, GRU, and LSTM models results in significantly larger prediction errors compared to RNN-7 and RNN-11 models, which incorporate these critical nuclear structure effects. Therefore, it is essential to highlight the best-performing model for comparison.

The variations in \( S_n \) and \( S_p \), as shown in Figures 5 and 6, highlight the impact of including shell and pairing terms in ML models. While most models show significant improvement when these effects are considered, small deviations remain, particularly in the GRU predictions for both \( S_n \) and \( S_p \). These deviations suggest the need for further optimization to fully capture fine nuclear structure details.

\begin{figure}
  \includegraphics[height=10cm]{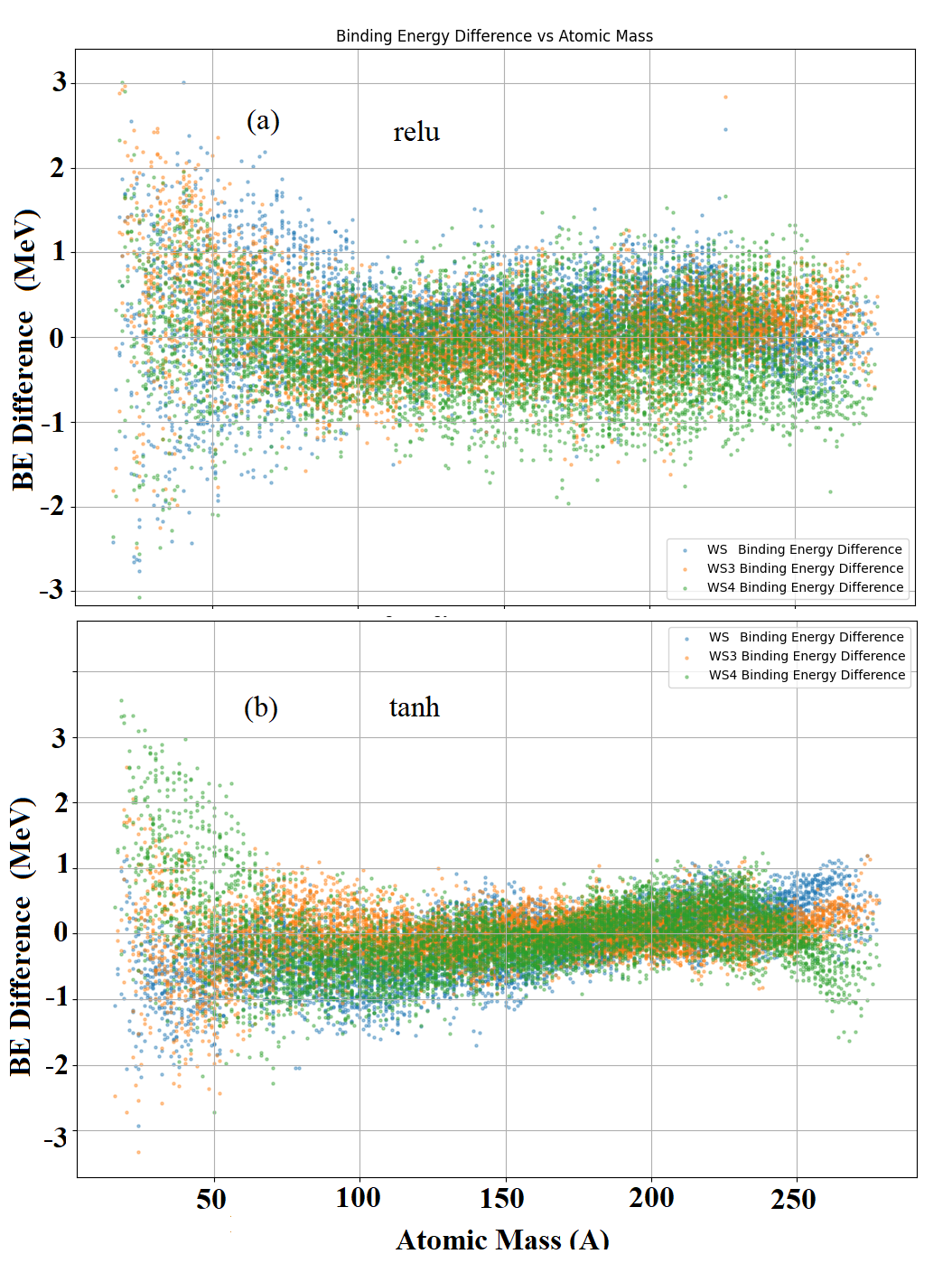}
\caption{Binding energy (BE) differences between GRU11 predictions using various activation functions and the WS, WS3, and WS4 models in the extrapolation region.}\label{relu}
\end{figure}

\begin{figure}
  \includegraphics[height=7cm]{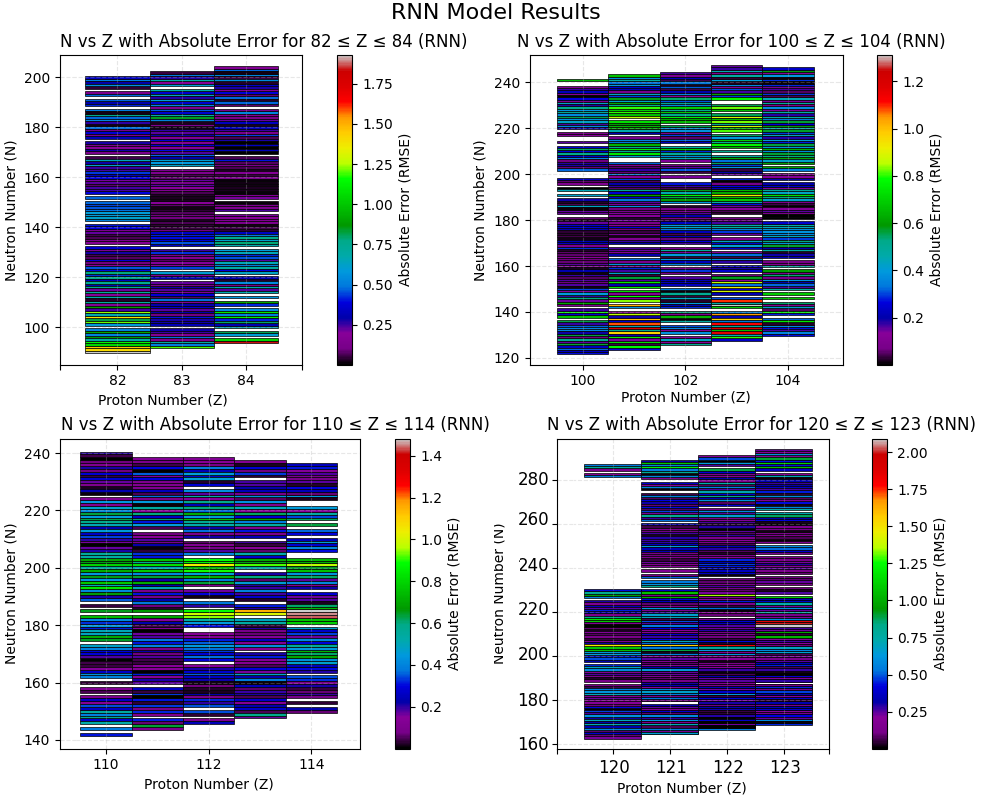}
\caption{Absolute $\sigma_{\mathrm{RMSE}}$ (in MeV) between WS3 and RNN11 predictions for different extrapolation regions: $82 \leq Z \leq 84$, $100 \leq Z \leq 104$, $110 \leq Z \leq 114$, and $120 \leq Z \leq 123$.}\label{rnn11}
\end{figure}
\begin{figure}
  \includegraphics[height=7cm]{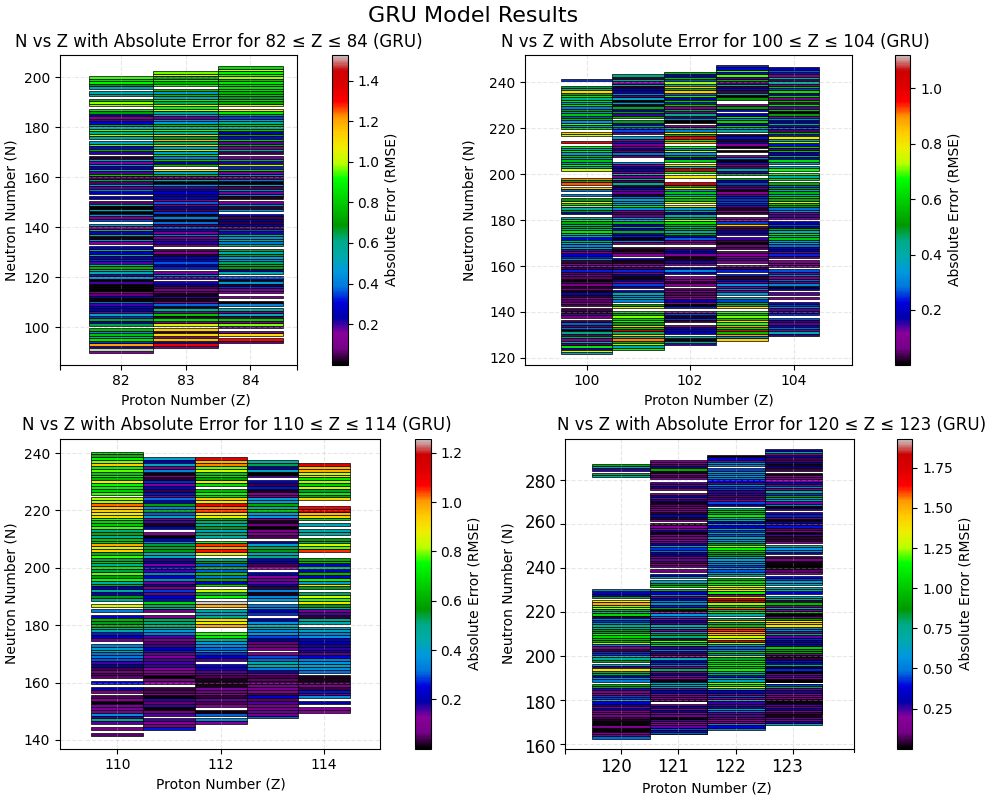}
  \caption{Absolute $\sigma_{\mathrm{RMSE}}$ (in MeV) between WS3 and GRU11 predictions for different extrapolation regions: $82 \leq Z \leq 84$, $100 \leq Z \leq 104$, $110 \leq Z \leq 114$, and $120 \leq Z \leq 123$.}\label{gru11}
\end{figure}
\begin{figure}
  \includegraphics[height=7cm]{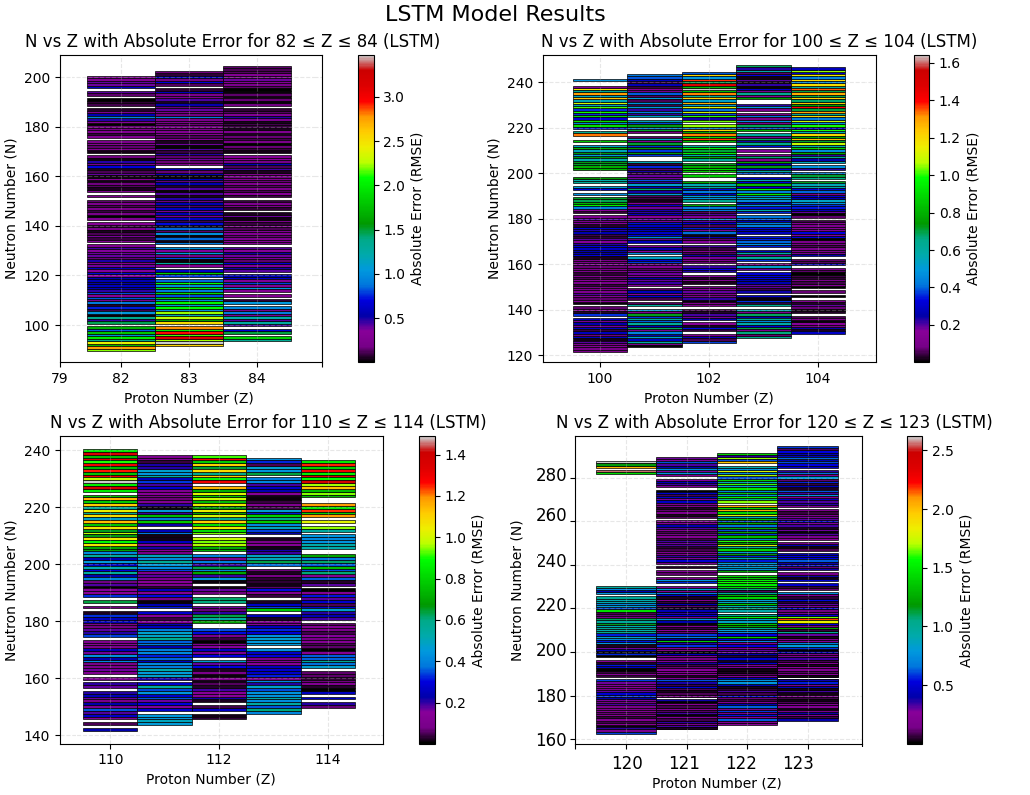}
  \caption{Absolute $\sigma_{\mathrm{RMSE}}$ (in MeV) between WS3 and LSTM11 predictions for different extrapolation regions: $82 \leq Z \leq 84$, $100 \leq Z \leq 104$, $110 \leq Z \leq 114$, and $120 \leq Z \leq 123$..}\label{lstm11}
\end{figure}

\subsection{Extrapolation Capabilities}

The evaluation of a model's extrapolation capabilities is essential in nuclear physics, particularly for predicting nuclear properties beyond experimentally known regions. In this section, we assess the extrapolation performance of the RNN-based models by examining their predictions for nuclear BEs near the neutron and proton drip lines.
The predictive power of the RNN-based models is benchmarked against established mass models, such as the WS4 model \cite{ws4}, which has demonstrated high accuracy in nuclear mass predictions. Several theoretical approaches, including the deformed Hartree-Fock-Bogoliubov method and nuclear density functional theory (DFT), have been widely employed to study ground-state and excited-state nuclear properties. Notably, nuclear DFT, despite utilizing a minimal set of parameters, has successfully predicted approximately 7,000 bound nuclides concerning neutron or proton emission, with this number potentially exceeding 10,000 when including continuum effects \cite{er,af,ag}.

In macroscopic-microscopic mass models, the WS4 model has achieved precise nuclear mass descriptions by fitting 18 parameters to 2,353 experimental mass data points, reaching a remarkable accuracy of 0.298 MeV \cite{ws4}. For comparison, previous studies have explored three widely used mass models-DZ, WS, and FRDM-alongside the liquid-drop model (LDM). To provide a clearer perspective, we compare the \( \sigma_\mathrm{RMSE} \) distributions of WS, WS3, and WS4 across the training, validation, and testing datasets for GRU, which emerged as the best-performing model among RNN variants.
A systematic analysis spanning $Z$ = 8 to $Z$ = 132, covering nuclei from the proton drip line to the neutron drip line, was conducted to evaluate the ground-state BE of these nuclei. Fig. ~\ref{relu} illustrates the BE differences between GRU-based model and extrapolation data, highlighting discrepancies in extrapolation predictions. These findings emphasize the necessity of larger datasets to enhance predictive performance. Notably, the WS3 macroscopic-microscopic model shows significant improvement with expanded datasets, nearly matching the performance of the GRU model.
The GRU model consistently demonstrates reliable and robust extrapolation capabilities, maintaining accurate predictions even in regions with sparse experimental data. The comparison between the \texttt{relu} and \texttt{tanh} activation functions for WS3 \cite{ws3} and GRU-calculated BEs reveals a substantial improvement when using \texttt{tanh}. Specifically, the deviation for extrapolation is reduced to 0.382 MeV, as detailed in Table VII. This improvement underscores the effectiveness of the GRU model in predicting BEs for unknown nuclei near the drip lines.
See Fig.~\ref{relu}.

\begin{table}[h]
\centering
\caption{\( \sigma_\mathrm{RMSE} \) for different models using \texttt{tanh} and \texttt{relu} activation functions. All (\(\sigma_\mathrm{RMSE}\)) are given in MeV.}
\begin{tabular}{cccc}
\hline
\hline
\multicolumn{4}{c}{\textbf{tanh AF}} \\
\textbf{Model} & \textbf{Training} & \textbf{Validation} & \textbf{Testing} \\
\hline
WS & 0.505 & 0.515 & 0.533 \\
WS3 & 0.382 & 0.401 & 0.406 \\
WS4 & 0.570 & 0.667 & 0.496 \\
\hline
\multicolumn{4}{c}{\textbf{relu AF}} \\
\textbf{Model} & \textbf{Training} & \textbf{Validation} & \textbf{Testing} \\
\hline
WS & 0.519 & 0.603 & 0.542 \\
WS3 & 0.477 & 0.533 & 0.478 \\
WS4 & 0.574 & 0.635 & 0.606 \\
\hline
\hline
\end{tabular}
\label{table:rmse_comparison}
\end{table}

To further illustrate these extrapolation capabilities, Figs.~\ref{rnn11},~\ref{gru11} and ~\ref{lstm11} present \( \sigma_\mathrm{RMSE} \) variations across different nuclear regions, specifically for $82 \leq Z \leq 84$, $100 \leq Z \leq 104$, $110 \leq Z \leq 114$, and $120 \leq Z \leq 123$ using the RNN11, GRU11, and LSTM11 models. These plots highlight the significant role of shell effects and pairing terms in improving nuclear mass predictions, with GRU consistently outperforming other models.


\section{Conclusion}\label{sec:conclusion}

In this study, we investigated the capability of RNN-based models, including GRU and LSTM, in predicting nuclear BEs and separation energies.
The augmentation of features led to a substantial improvement in $\sigma_\mathrm{RMS}$ between RNN-based predictions and the AME2020 data.
Specifically, the inclusion of bulk properties, surface terms, and Coulomb contributions initially reduced the error from several MeV to sub-MeV levels. Further refinement by incorporating nuclear shell effects and pairing terms led to an even more significant reduction, bringing deviations down to the hundred-keV scale in the GRU model. Our results demonstrate that including these nuclear structure effects substantially enhances predictive performance, minimizing $\sigma_\mathrm{RMS}$  deviations compared to models that neglect them.
A comparative analysis with established mass models such as WS3 highlights the reliability of DL approaches in nuclear mass predictions. The GRU model, in particular, exhibited superior generalization capabilities, achieving competitive \( \sigma_\mathrm{RMSE} \) values comparable to those of macroscopic-microscopic models while maintaining robustness across different nuclear regions. The systematic extrapolation study, spanning from $Z$ = 8 to $Z$ = 132, further confirmed the model's effectiveness in predicting nuclear masses near the neutron and proton drip lines.
Additionally, we evaluated the influence of activation functions on model performance. The results indicate that using the \texttt{tanh} activation function yields lower \( \sigma_\mathrm{RMSE} \) values compared to \texttt{relu}, particularly in WS3 and WS4 mass models. This suggests that \texttt{tanh} activation better captures the complex correlations in nuclear structure, enhancing predictive accuracy.
The findings underscore the potential of DL techniques in nuclear physics, particularly in regions where experimental data is sparse or unavailable. While the GRU-based model provides a promising approach to nuclear mass predictions, further improvements, such as incorporating additional physical constraints or hybridizing DL with macroscopic-microscopic models, could further enhance predictive accuracy.
Future work will focus on extending these models to other nuclear observables, refining hyperparameter tuning strategies, and integrating uncertainty quantification to assess the reliability of extrapolations. The integration of physics-informed NN and transfer learning could also provide new avenues for improving model generalizability across the nuclear landscape.


\begin{acknowledgements}\scalebox{0.9}
This work was supported by the National Natural Science Foundation of China (Grant No. 12250410254 and No. 12175199) , the ZSTU intramural grant (Grant No. 23062211-Y).
\end{acknowledgements}

\end{document}